 \documentclass[12pt]{article}
  \usepackage{youngtab}
 \usepackage{url}
  
\catcode`@=11
\def\secteqno{\@addtoreset{equation}{section}%
\def\theequation{\thesection.\arabic{equation}}}
\catcode`@=12 \secteqno
\newcommand{\be}{\begin{equation}}
\newcommand{\ee}{\end{equation}}
\newcommand{\bea}{\begin{eqnarray}}
\newcommand{\eea}{\end{eqnarray}}
\newcommand{\bref}[1]{(\ref{#1})}
\newcommand{\nn}{\nonumber}

\newcommand{\ddx}[1]{\frac{\partial}{\partial #1}}

\def\repnum{PoincareExtensions.tex (\today)}
\newcommand{\bi}{\begin{enumerate}}
\newcommand{\ei}{\end{enumerate}}
  
 \newcommand{\D}{\delta} 
 
\newcommand{\T}{\theta}

\newcommand{\h}{\eta}

\def\6{\partial}
\def\7{\tilde}
\def\8{\hat}

\def\CL{{\cal L}}

\def\={{\;=\;}}\def\+{{\;+\;}}

\begin{document}
\thispagestyle{empty} \hfill\today

\hfill ICCUB-08-147, KEK-1292, UB-ECM-PF-08/23

\vskip 20mm
\begin{center}
{\Large\bf

Infinite Sequence of Poincar{\'e} Group Extensions: Structure and
Dynamics} \vskip 10mm {\large ~ Sotirios Bonanos$^1$
and Joaquim Gomis$^{2,3}$} \vskip 6mm
\medskip

$^1$Institute of Nuclear Physics, NCSR Demokritos,\\
15310 Aghia Paraskevi, Attiki, Greece\\
\vspace{6pt}  $^2$Departament d'Estructura i Constituents de la
Mat\`eria and ICCUB, Universitat de Barcelona, Diagonal 647, 08028 Barcelona\\
\vspace{6pt}
$^3$High Energy Accelerator Research Organization (KEK),\\
 Tsukuba, Ibaraki, 305-0801 Japan \vskip 6mm
 {\small e-mail:\
 {sbonano@inp.demokritos.gr\\gomis@ecm.ub.es\\} }

\medskip

\end{center}

\vskip 20mm
\begin{abstract}
We study the structure and dynamics of the infinite sequence of
extensions of the Poincar{\'e} algebra whose  method of construction
was described in a previous paper \cite{Bonanos:2008kr}. We give
explicitly the Maurer-Cartan (MC) 1-forms of the extended Lie
algebras up to level three. Using these forms  and introducing a corresponding set of new dynamical couplings, we construct an invariant Lagrangian, which describes the dynamics of a distribution of charged particles in an external electromagnetic field. At each extension, the distribution is approximated by a set of moments about the world line of its center of mass and the field by its Taylor series expansion about the same line. The equations of motion after the second extensions contain back-reaction terms of the moments on the world line.

\end{abstract}

\vskip 4mm

\setcounter{page}{1}
\parskip=7pt
\newpage
\section{Introduction}

In a recent  paper \cite{Bonanos:2008kr} we have studied aspects of
the Chevalley-Eilenberg cohomology of the Galilei and Poincar{\'e}
groups. In particular we have seen that, at degree two, there is an
infinite sequence of Lie algebra extensions, beginning with
 the Galilei or Poincar{\'e} algebras. We recall
that the extensions found are non-central, since the corresponding
generators transform non-trivially under the corresponding  normal
subalgebra of the unextended algebra.

In this paper we study further the infinite sequence of extensions
of the Poincar{\'e} algebra.  We study the tensor structure of the
extensions and their physical interpretations.
 It is known that  the Poincar{\'e} group  in d+1 ($d>1$) has no
central extensions\footnote{In $1+1$ dimensions  there is one central extension
\cite{levyleblond69}, which has been used to
study several problems of gravity and Moyal quantization, see for
example \cite{Aldaya:1985va,olmo1,Cangemi:1992bj}. Also,  in spaces $R^{2n}$ and $R^{4n}$ with automorphism groups $U(n)$ and $Sp(n)\times SU(2)$ (K\"{a}hler and hyper-K\"{a}hler geometries),
Galperin et.al. \cite{Galperin:1987wc,Galperin:1987wb} have obtained
complex first-level central extensions (a triplet in hyper-K\"{a}hler).}
 and that it has a non-central antisymmetric
tensor extension \cite{galindo}.
Physically, this extension
corresponds to the symmetries of a relativistic particle in a
constant electromagnetic field and is known  as  the Maxwell group
 \cite{Schrader:1972zd} \cite{Beckers:1983gp}.  A modification of the Poincar{\'e} algebra having only two Lorentz generators, that leaves a constant background electromagnetic field invariant, allows an extension with two central charges that can be interpreted as the electric and magnetic charge  (BCR algebra \cite{Bacry:1970ye}). Non-central extensions have also been considered for the diffeomorphism gauge algebra, see for example \cite{larsson}.

Both central and non-central extensions are controlled by
Chevalley-Eilenberg  cohomology theory, see for example
\cite{azcarragabook}.
Here we compute in a systematic, almost algorithmic\footnote{The
calculations  make use of the first author's Mathematica package EDC
(Exterior Differential Calculus) \cite{bonanos}. The procedure is described in detail in  \cite{Bonanos:2008kr}.},  way the most
general CE cohomology groups at form degree  two.
As we will see the non-trivial forms belong to different
representations of the subgroup of Lorentz transformations of the
Poincar{\'e}  group, in general with mixed symmetries.  The  Lorentz
transformations are a subgroup  of the automorphism group and
constitute a normal subgroup of each
extended group. We can associate to any non-central extension a
Young tableau.

As discussed in  \cite{Bonanos:2008kr}, the first  non-central extension of the   Poincar{\'e} algebra is
obtained by calculating  all possible non-trivial closed 2-forms of
the subgroup of  space-time translations.  The forms are closed with
respect the exterior differential operator $d$. The complete
extended algebra is constructed from the original algebra and the
extensions by incorporating their transformation properties  under Lorentz
transformations, which is equivalent to replacing the exterior
 differential operator $d$ by the corresponding  ``covariant"
 operator $d+M \wedge$, $M$ being the zero-curvature connection
associated to the Lorentz generators: $dM+M\wedge M=0$.

Once we have an extended algebra, we can further extend it by
applying the same procedure to   an extended set of ``translations"
which includes all generators except those of the Lorentz subgroup.
In this way we obtain new extensions whose generators belong to
higher dimensional representations of the  Lorentz group.  This
procedure does not end resulting in an infinite sequence of groups -- extensions
of the Poincar{\'e} group. In the limit, our procedure formally defines an infinite Lie algebra. 
However, we cannot prove this result. 

We do not have a precise mathematical interpretation for this
infinite Lie algebra.
 We find it intriguing, however, that the generator content of this algebra is
organized in levels like the Lorentzian Kac-Moody algebras that are
conjectured to be a symmetry of supergravity; see, for example,  \cite{West:2001as} for the $E_{11}$
approach and \cite{Damour:2002cu} for the $E_{10}$ approach.

In order to obtain a possible physical interpretation
of this infinite
sequence of extensions of the Poincar{\'e} group we construct a
relativistic particle Lagrangian, invariant  under the extended
algebra, by using the MC forms. We also
introduce tensor coupling ``constants" that we consider as  new dynamical
variables. These tensor couplings are invariant under the extended
symmetries.

 The form of the equations of motion following from this Lagrangian leads us to the conclusion that the physical system in question is a distribution of charged particles, described collectively as a particle with a  set of multipole moments, moving in a fixed background electromagnetic field. The multipoles can be considered as Goldstone bosons.
The background field is described by its Taylor series expansion about the world line of the ``particle", higher terms in the series (and higher moments) appearing with every extension.
Moreover, new terms in the equation of motion of the ``particle" due to back-reaction terms involving the moments appear.
These results are  obtained
by integrating  the equations of motion for the  coupling   fields
and plugging the solutions into the  equations of motion of
the particle coordinates.  Once we choose  a  particular solution,
the equations of motion for the particle coordinates imply a
spontaneous breaking of the symmetries of the extended algebra.

The organization of the paper is as follows. In section 2 we introduce
our notation and conventions and  obtain the first level extensions.
We then find explicit expressions for the generators of the extended group,
 construct the Lagrangian, and deduce the transformations of the fields that
 leave the Lagrangian invariant. Finally, we obtain the equations of motion for
 all dynamical variables. In sections 3 and 4 we repeat these steps for the second
 and third extensions. We also give the defining equations for the fourth level extensions.
 In section 5 we point out that it is advantageous to consider the Young Tableau symmetries
 of the different  extensions; and how symmetry considerations can determine the structure of
 higher extensions.
Finally, in section 6 we compare our results with other approaches
for constructing theories with higher symmetry and discuss the
implications.

\section{The Poincar{\'e} group in 3+1 dimensions}
The generators of the unextended Poincar{\'e} algebra  are the translations
$P_{a}$ and the Lorentz transformations  $M_{ab}$, where the tensor
 indices  take the values $(0,\,1,\,2,\,3)$. Denoting by $\h_{ab}$
 the Minkowski metric, the algebra is given by\footnote{ Although the algebra
  is real and the imaginary units
can be made to disappear by replacing all generators $G$ by $i G'$,
we prefer to leave the $i$'s in the equations because then we can
interpret the generators as Hermitian operators.}
\bea
\left[M_{ab},~M_{cd}\right]&=&-i~(\h_{bc}~M_{ad}-\h_{bd}~M_{ac}+\h_{ad}~M_{bc}-\h_{ac}~M_{bd}),\nn\\
\left[P_a,~M_{bc}\right]&=&-i~(\h_{ab}~P_{c}-\h_{ac}~P_{b}).\label{Poincare0} \eea

 As described in  \cite{Bonanos:2008kr}, in order to construct the extensions we   make use of
  the left invariant
Maurer-Cartan (MC) form, defined by \be \label{omega}\Omega=-i
g^{-1}dg,\ee where $g$ represents a general element of the Poincar{\'e}
group. The MC form satisfies the Maurer-Cartan equation \be
d\Omega+i~\Omega\wedge\Omega=0.\label{MCeq}\ee In components the MC
1-form is written, for a generic Lie algebra, as \be
\label{MC1}\Omega= X_A~  {\cal X}^A,\ee where $X_A$
are the generators of the Lie algebra satisfying  \be
[X_B,~X_C]=i~{f^A}_{BC}~X_A \label{CommDef} \ee and $ {\cal X}^A$ are corresponding
1-forms.\footnote{In general the generator indices will refer to multiple-index tensors with symmetries. When such indices are summed as in \bref{MC1}, \bref{CommDef},  additional numerical factors must be introduced to compensate for multiple appearances of identical terms in the sum.}
Throughout this paper we use the same capital letters in plain and
calligraphic font  to denote generators and associated 1-forms. The
MC equation  \bref{MCeq}  implies that the 1-forms  $ {\cal X}^A$
satisfy \be d{\cal X}^A=\frac12{f^A}_{BC} {\cal X}^B\wedge {\cal
X}^C. \ee

For the Poincar{\'e} case, the MC 1-form \bref{MC1} becomes \be \Omega=
P_a~{\cal P}^a+\frac12 M_{ab}~{\cal M}^{ab},\ee
while the MC equation  \bref{MCeq} in components is
\bea
d{\cal P}^a&+&{\cal P}^c \wedge{{\cal M}_c}^a=0,
\nn\\
d{\cal M}^{ab}&+&{\cal M}^{ac} \wedge{{\cal M}_c}^b\;=0.
\label{MCNH} \eea

  The first step in the cohomological analysis is to
 freeze the Lorentz degrees of freedom   and construct the most general
 2-form that can be built from the  translations alone,   ${\cal P}^a$. The MC equations
for these generators ($d{\cal P}^a=0$) are  obtained by putting
${\cal M}^{ab}\rightarrow 0$ in \bref{MCNH}.  Then we find that the
most general closed invariant 2-form which cannot be written as the
differential of an invariant 1-form is

\be\label{general2form} \Omega_2= f_{\left[ab \right]}\,{\cal
P}^a\wedge{\cal P}^b\ee where  $f_{\left[ab \right]}$ is a constant
second rank antisymmetric tensor.
Therefore  the non-trivial 2-form  extensions  belong to  an
antisymmetric tensor representation of the Lorentz group. The 1-form
``potentials" associated to these closed 2-forms are denoted by ${\cal
Z}^{\left[ab \right]},$ and are defined by the equation
 \be d{\cal Z}^{\left[ab
\right]}={\cal P}^a\wedge{\cal P}^b \label{dZeqsM0}. \ee

From this equation we obtain the algebra of  the corresponding
generators, denoted by
 $Z_{\left[ab \right]}$. We find
 \be \left[P_{a},~P_{b}\right]= +i~Z_{\left[ab\right]},  \label{NHm1} \ee
which implies that there is no central extension of
the Poincar{\'e} group.

With the rotations included, the extended set of MC 1-forms
satisfies the equations
 \bea
 d{\cal P}^a&=&-{\cal P}^c \wedge{{\cal M}_c}^a,
\nn\\
d{\cal M}^{ab}&=&-{\cal M}^{ac} \wedge{{\cal M}_c}^b,\label{extMCNH}\\
d{\cal Z}^{\left[ab \right]}&=&-{\cal Z}^{\left[ac \right]} \wedge {{\cal M}_c}^b-
{{\cal M}^a}_c \wedge{\cal Z}^{\left[cb \right]} +{\cal P}^a \wedge{\cal P}^b. \nn
\eea
The associated algebra was introduced before in the literature
\cite{Bacry:1970ye,Schrader:1972zd,Beckers:1983gp}. It is known as the
Maxwell algebra.

\subsection{Explicit parametrization}

Here we will introduce explicit parameters labeling   the group elements, which will induce an explicit parametrization of all MC 1-forms in terms of the differentials of these parameters. We will first obtain expressions for the MC 1-forms without  the Lorentz generators.
Specifically, the  general element of this coset of the extended group will be parametrized, locally, by
\be\label{parametrization2}
g=e^{iP_ax^a}e^{\frac{i}{2}Z_{ab}\theta^{ab}}. \ee
where $x^a, ~\theta^{[ab]}$ are  the group parameters associated to the generators $P_a,~Z_{ab}$.
The component MC 1-forms \bref{MC1} can be computed directly from the definition \bref{omega} and the commutator \bref{NHm1} using the
Baker-Campbell-Hausdorff formula. The result is
\be {\cal P}^a=dx^a, \hspace{30 pt} {\cal Z}^{\left[ab\right]}=d\theta^{\left[ab \right]}+\frac12(x^a\;dx^b-x^b\;dx^a).
\label{explicit1} \ee

 If we want to have the explicit
expressions of these MC 1-forms when we include the Lorentz degrees of freedom, the   right
hand side of all vector and tensor expressions given above must be
 multiplied by an appropriate   orthogonal matrix, $U$, for each index:
\be {\cal P}^a={{U^{-1}}^a}_b\;dx^b,\hspace{30 pt} {\cal
Z}^{[ab]}= {{U^{-1}}^a}_p\;{{U^{-1}}^b}_q\;
\left(d\theta^{[pq]}+\frac12(x^p\;dx^q-x^q\;dx^p)\right).\ee
$U$ depends on the parameters associated to the Lorentz generators and can be obtained by adding appropriate terms to \bref{parametrization2}.
In terms of $U$ the  Lorentz  MC 1-forms   have the explicit representation ${{\cal M}^{a}}_{b}= {{U^{-1}}^a}_c\; d{U^c}_b$.

The vectors fields dual to these MC forms (when we freeze the Lorentz degrees of freedom) are:
 \bea
Z_{ab}&=&-i~\ddx{\theta^{ab}},\\
P_a&=&-i~(\ddx {x^a}+\frac{1}{2} x^r \ddx{\theta^{ar}}),\\
M_{ab}&=&-i~(x_a \ddx {x^b}-x_b \ddx {x^a}+{\theta_a}^{r}
\ddx {\theta^{br}}-{\theta_b}^{r} \ddx {\theta^{ar}}),
\eea
and satisfy the commutators
\bea
\left[M_{ab},~M_{cd}\right]&=&-i~(\h_{bc}~M_{ad}-\h_{bd}~M_{ac}+\h_{ad}~M_{bc}-\h_{ac}~M_{bd}),\nn\\
\left[P_a,~M_{bc}\right]&=&-i~(\h_{ab}~P_{c}-\h_{ac}~P_{b}),\nn \\
\left[P_a,~P_b\right]&=&+i~Z_{ab},\\
\left[M_{ab},~Z_{cd}\right]&=&-i~(\h_{bc}~Z_{ad}-\h_{bd}~Z_{ac}+\h_{ad}~Z_{bc}-\h_{ac}~Z_{bd}),\nn\\
\left[P_a,~Z_{bc}\right]&=&~0,\nn\\
\left[Z_{ab},~Z_{cd}\right]&=&~0.\nn
 \eea

\subsection{Particle Lagrangian and Noether charges}

To obtain a physical interpretation,  one possibility is to
construct a particle Lagrangian that is invariant under the extended
Poincar{\'e} group using the non-linear realization method
\cite{Coleman} for space-time symmetries, see for example
\cite{Gomis:2006xw}. A diffeomorphism-invariant  free particle
Lagrangian is $\CL_0=m\sqrt{-\dot x_a^2}$ and depends on the
translation 1-forms only. A possible Lagrangian including also the
first extension 1-forms ${\cal Z}^{ab}$ \bref{explicit1}, is \be
\label{lagfirst} \CL=m\sqrt{-\dot x_a^2}+\frac12f_{ab}{\cal
Z}^{ab}=m\sqrt{-\dot x_a^2}+\frac12 f_{ab}(\dot\T^{ab}+x^{[a}\dot
x^{b]}), \ee where we have introduced  the  antisymmetric  tensor
couplings $f_{ab}(\tau)$ that are considered as new dynamical
variables, in addition to the group space coordinates ($x^a,
\T^{ab}$), which are now also functions of the particle proper time.
The physical interpretation of the extra variables $\T^{ab}, f_{ab}$
will be given after the equations of motion have been obtained.

This Lagrangian is invariant under
translations\footnote{The generators  of these transformations
are the right invariant vector fields.} \bea \D_P
x^{a}&=&\epsilon^a,\\
\D_P \T^{ab}&=& -\frac12(\epsilon^a x^b-\epsilon^b x^a),\\
\D_P f_{ab}&=&0,
 \eea and the non-vanishing shifts
\be \D_Z \T^{ab}=\epsilon^{ab}, \quad \D_Z f_{ab}=0.\ee

The Noether charges associated to these symmetries are \be
P_a=p_a-\frac12p_{ab}x^b,\quad\quad Z_{ab}=p_{ab}.\ee If we compute the Poisson
bracket among these generators we find  \be
\{P_a,P_b\}=-Z_{ab},\ee
i.e., we recover the algebra\footnote{The reason for the overall sign
difference from the starting algebra  is that now the generators are
{\it active} operators.} \bref{NHm1}.

 In  the proper time gauge the Euler-Lagrange equations of
motion following from \bref{lagfirst} are
\bea\label{eqmotion1theta} \D
\T^{ab}&\to& \dot f_{ab}=0,
\\\label{eqmotion1f} \D f_{ab}&\to& \dot
\T^{ab}+\frac12(x^a\dot x^b-x^b\dot x^a)=0,
\\\label{eqmotion1x}
\D x^{a}&\to& -m\ddot x_a+f_{ab}\dot x^b=0.\eea

Equation \bref{eqmotion1x}, with $f_{ab}=e\,F_{ab}$, is the Lorentz
force equation determining the motion of a particle of mass $m$ and
charge $e$ in an electromagnetic field $F_{ab}$.  Note that for this
case the equation of motion for $f_{ab}$, \bref{eqmotion1f} does not
affect the dynamics of the coordinates.
This equation tells us that
$\dot\theta^{ab}$ is proportional to
 the $ab$ component of the
angular momentum (or  magnetic moment) of the  particle.\footnote{The terminology refers to the space-space components; the space-time components give a Lorentz-boosted momentum (or dipole moment).}
 In other words $\theta^{ab}$ is a non-local function  of
the components of the angular momenta of the particle.

Integration of the equation of motion associated to $\theta$ gives
$f_{ab}=f_{ab}^0=e\,F_{ab}^0$. We see that this solution spontaneously breaks
Lorentz symmetry.
If we substitute this solution in the equation of motion
 for the variable $x$,  \bref{eqmotion1x},
we find  that it describes the motion of a particle in a constant,
fixed EM field with \be F_{ab}=F_{ab}^0=constant. \ee It can be
obtained from the potential \bea A_a=-\frac12F_{ab}^0x^b. \eea

\section{Second  level extensions}

One can obtain further extensions of the Poincar{\'e} group which lead to
new generators in higher dimensional representations of the
Lorentz group. In order to find them we apply  the same procedure as in the last
section, at every level  taking as  ``translations" all generators of the previous level other
than the  Lorentz  ones, ${\cal M}^{ab}$.

 For the second extension we take as  ``translations"  the 1-forms \be {\cal P}^a,~ {\cal Z}^{\left[ab \right]}.\ee
The calculation  results in 20 closed
non-trivial 2-forms which can be written as the components  the
tensor\footnote{This tensor is antisymmetric in $\left[bc\right]$
and its totally antisymmetric part vanishes. This leads to 4
identities, $\epsilon_{abcd} {{\cal P}^b} \wedge{\cal Z}^{\left[cd
\right]}=0$, leaving 4x6-4=20 independent components.}
 \be 2~{{\cal P}^a} \wedge{\cal Z}^{\left[bc \right]} - {{\cal
P}^b} \wedge{\cal Z}^{\left[ca \right]}-{{\cal P}^c}\wedge{\cal
Z}^{\left[ab \right]}. \ee Again, introducing the second-level
potential 1-form ${\cal Y}^{a\left[bc\right]}$, with the same
symmetries as the above 2-form tensor and unfreezing the Lorentz
freedom, we find \bea d{\cal Y}^{a\left[bc\right]}&=&-{{\cal M}^a}_s
\wedge{\cal Y}^{s\left[bc\right]} -{{\cal M}^b}_s \wedge{\cal
Y}^{a\left[sc\right]}-{{\cal M}^c}_s \wedge{\cal
Y}^{a\left[bs\right]} \nn \\& & +~2~{{\cal P}^a} \wedge{\cal
Z}^{\left[bc \right]}  - {{\cal P}^b} \wedge{\cal Z}^{\left[ca
\right]}-{{\cal P}^c}\wedge{\cal Z}^{\left[ab \right]}.\label{dYeqs} \eea
 The corresponding generators $Y_{a[bc]}$ appear in the commutators of
  the original translations with the first level extensions
  \be\label{step2}
 \left[P_a,~Z_{\left[bc \right]}  \right]=2~i~Y_{a\left[bc\right]}
 -i~Y_{b\left[ca\right]}-i~Y_{c\left[ab\right]}.\ee

 The complete set of  the second extension commutators are:
\bea
\left[M_{ab},~M_{cd}\right]&=&-i~(\h_{bc}~M_{ad}-\h_{bd}~M_{ac}+\h_{ad}~M_{bc}-\h_{ac}~M_{bd}),\nn\\
\left[P_a,~M_{bc}\right]&=&-i~(\h_{ab}~P_{c}-\h_{ac}~P_{b}),\nn \\
\left[P_a,~P_b\right]&=&i~Z_{ab},\nn\\
\left[M_{ab},~Z_{cd}\right]&=&-i~(\h_{bc}~Z_{ad}-\h_{bd}~Z_{ac}+\h_{ad}~Z_{bc}-\h_{ac}~Z_{bd}),\nn \\
\left[P_a,~Z_{bc}\right]&=&~i~(2~Y_{abc}-Y_{bca}-Y_{cab}),\label{algebra2}\\
\left[Z_{ab},~Z_{cd}\right]&=&~0,\nn\\
\left[Y_{pab},~M_{cd}\right]&=&-i~(\h_{bc}~Y_{pad}-\h_{bd}~Y_{pac}+\h_{ad}~Y_{pbc}-\h_{ac}~Y_{pbd}+\h_{pc}~Y_{dab}-\h_{pd}~Y_{cab}),\nn\\
\left[Y_{pab},~Z_{cd}\right]&=&~0,\nn\\
\left[Y_{pab},~P_{c}\right]&=&~0,\nn\\
\left[Y_{pab},~Y_{qcd}\right]&=&~0.\nn
 \eea
Note that, at this level,  the operators  $Z_{ab}$ and  $Y_{abc}$
generate an Abelian subgroup.

\subsection{Explicit parametrization}
Introducing the second extension parameters $\xi^{a[bc]}$
(coordinates in group space) with the symmetries of ${\cal
Y}^{a[bc]}$, the coset element in the second extension can be
written as \be
g=e^{iP_ax^a}e^{\frac{i}{2}Z_{ab}\theta^{ab}}e^{\frac{i}{2}Y_{abc}\xi^{abc}}.
\ee We can then compute, as before, the corresponding MC forms; the
ones associated to the translations and first extension are not
modified, while the second extension MC forms are found to be \be
{\cal Y}^{abc}=d\xi^{abc}-2dx^a\T^{bc}+dx^b\T^{ca}+dx^c\T^{ab}+
\frac12x^{a}(x^{b}dx^c-x^{c}dx^b).\ee

  The differential operators dual to the 1-forms given above provide a representation of the extended algebra \bref{algebra2} (in the summations below, differentiations with respect to variables that are zero -- $\theta^{00}$, $\xi^{011}$, etc., --  are omitted):
 \bea
 Y_{abc}&=&-i~\ddx{\xi^{abc}},\\
Z_{ab}&=&-i~\ddx{\theta^{ab}},\\
P_a&=&-i~(\ddx {x^a}+\frac{1}{2} x^r \ddx{\theta^{ar}}+\theta^{rs} \ddx{\xi^{ars}}-
\theta^{rs} \ddx{\xi^{rsa}}-\frac12x^r x^s \ddx{\xi^{rsa}}),\\
M_{ab}&=&-i~[x_a \ddx {x^b}-x_b \ddx {x^a}+{\theta_a}^{r} \ddx
{\theta^{br}}-{\theta_b}^{r} \ddx {\theta^{ar}}\nn
\\&+&\frac12({\xi_{a}}^{rs} \ddx{\xi^{brs}}-{\xi_{b}}^{rs}
\ddx{\xi^{ars}})+{\xi^{rs}}_{a} \ddx{\xi^{rsb}}-{\xi^{rs}}_{b}
\ddx{\xi^{rsa}}]. \eea

\subsection{Lagrangian associated to the second level extension}
When we include the second extension, the particle Lagrangian becomes
 \be \CL= m\sqrt{-\dot x_a^2}+\frac12 f_{ab}{\cal Z}^{ab}+\frac12 f_{abc}{\cal Y}^{abc},\ee
where the new  tensor couplings $f_{abc}(\tau)$  have the symmetries
of ${\cal Y}^{abc}$ and,  together with the new group space
coordinates $\xi^{abc}(\tau)$, are considered as new dynamical
variables.

Apart from the ordinary Lorentz transformations, the non-trivial
transformations  leaving the Lagrangian invariant  are
\bea \D_P
x^{a}&=&\epsilon^a,\\
\D_P \T^{ab}&=&- \frac 12 (\epsilon^{a} x^{b}-\epsilon^{b} x^{a}),\\
\D_P \xi^{abc}&=& -x^a(\epsilon^{b} x^{c}-\epsilon^{c} x^{b}).
 \eea
\bea
\D_Z \T^{ab}&=&\epsilon^{ab},\\
\D_Z \xi^{abc}&=&  2 x^a\epsilon^{bc}- x^b\epsilon^{ca}- x^c\epsilon^{ab},
 \eea
\bea \D_Y \xi^{abc}= \epsilon^{abc},
 \eea
 where $\epsilon^{a},~\epsilon^{ab},~\epsilon^{abc}$ are arbitrary displacements of the group parameters.
 The conserved quantities are written as \be
Q=\epsilon^a P_a+\frac
12\epsilon^{ab}Z_{ab}+\frac12\epsilon^{abc}Y_{abc}, \ee from which
we obtain the Noether generators \bea
P_a=p_a-\frac12p_{ab}x^b+p_{bca}x^bx^c, \quad
Z_{ab}=p_{ab}+x^c(2p_{cab}-p_{bca}-p_{abc}), \quad Y_{abc}=p_{abc}.
\eea The Poisson brackets among these generators reproduce the
algebra \bref{algebra2} up to a sign.

The Euler-Lagrange  equations of motion now take the form
\bea
\D \xi^{abc}&\to& \dot f_{abc}=0,\label{theta3}\\
\D \T^{ab}&\to& \dot f_{ab}=(-2f_{cab}+f_{abc}+f_{bca})\dot x^c,\label{theta2}\\
\D f_{abc}&\to& \dot \xi^{abc}-2\dot x^a\T^{bc}+\dot
x^{b}\T^{ca}+\dot x^{c}\T^{ab}+
\frac12x^{a}(x^{b}\dot x^{c}-x^{c}\dot x^{b})=0, \label{eomf3} \\
\D f_{ab}&\to& \dot \T^{ab}+\frac12(x^{a}\dot x^{b}-x^{b}\dot
x^{a})=0,
\label{eomf2}\\
\D x^{a}&\to& -m\ddot x_a+f_{ab}\dot x^b=-\frac12 \dot f_{ab}x^b+\frac12(-2f_{abc}+f_{bca}+f_{cab})
\dot \T^{bc}\nn \\
&  &-\frac12(-2f_{bca}+f_{cab}+f_{abc})x^b \dot x^{c}
\label{eomx} .\eea

Substituting for $\dot f^{ab}, ~\dot \T^{ab}$  from \bref{theta2}, \bref{eomf2},
we find that the RHS of   \bref{eomx} vanishes, so that the
 equation of motion for $x^a$ depends only on $f_{ab}$:
 \be-m\ddot x_a+f_{ab}\dot x^b=0.\label{eomx0rhs}\ee

 If we integrate equations  \bref{theta3}, \bref{theta2} for $f_{abc}$ and $f_{ab}$, we
get \be  f_{abc}=f^0_{abc},
 \quad  \quad f_{ab}=(-2f^0_{cab}+f^0_{abc}+f^0_{bca}) x^c+f^0_{ab},\quad
 \quad f^0_{\dots}=const.\label{fab2} \ee Note that these solutions break  the extended symmetry spontaneously.
If we substitute these expressions back in \bref{eomx0rhs} we have

\be -m\ddot x_a+ f^0_{ab}\dot x^b+(-2f^0_{cab}+f^0_{abc}+f^0_{bca})
x^c \dot x^b=0.\ee The resulting equation of motion describes a
particle in a  given  EM field which is linear in the cartesian
coordinates and can be obtained from the potential \bea
A_a=F^0_{cab}x^bx^c-\frac12F^0_{ab}x^b, \quad \quad f^0_{cab}=e\,F^0_{cab}.\label{Apot2} \eea

What makes this possible is the symmetry properties of the
quantities $f^0_{cab}$ which imply that the field strength $f_{ab}$
\bref{fab2} satisfies $f_{[ab,c]}=0$.

Note that, as in the previous level, the equation of motion for
$f_{ab}$ and $f_{abc}$ do not affect the dynamics of the
coordinates. The variable $\theta^{ab}$ retains its old
interpretation in terms of the magnetic moment of the particle,
while, from \bref{eomf3} with $\theta^{ab}=0$, we see that
$\xi^{abc}$ is related to the integral of the first moment of the
magnetic moment, i.e.,  the magnetic quadrupole moment (second
moment of the current distribution). Thus, it appears that our
physical system is a {\it distribution} of charged particles,
described collectively   at this level  as the motion of a particle with   two sets
of moments $\theta^{ab}, \xi^{abc}$, moving in a given EM field.
The non-locality of the
equations determining the moments suggests that the particles also
interact among themselves. This will become apparent at the next
level where the equation determining $x^a$ will acquire an extra
force term proportional to the magnetic moment.

Writing \bref{theta2} as $df_{ab}=(-2f_{cab}+f_{abc}+f_{bca})d x^c$,
we can interpret the coefficients $f_{abc}$ as giving the partial
derivatives of $f_{ab}$.

\section{Higher level extensions}
In this section we will consider explicitly the higher
order extensions up to level four. Here, as we will see, a new
phenomenon appears:  we need more than one tensor to describe the new extensions.
Moreover, some of the lower level  extensions no longer
commute with themselves or with the translations.

The procedure can be continued indefinitely. It will become apparent that, at
 level $n$, several new tensor extensions of rank $n+1$ appear.

\subsection{Third extension}
At the third level the procedure gives 60 closed 2-forms which can be arranged as
 the components of two 4th rank tensors with the following symmetries:
${\cal S}_1^{(ab)(cd)}, ~~ {\cal S}_2^{[ab][cd]}$ and the additional antisymmetry
${\cal S}_1^{(ab)(cd)}=-{\cal S}_1^{(cd)(ab)}$ and  ${\cal S}_2^{[ab][cd]}=-{\cal S}_2^{[cd][ab]}$.
These 1-form potentials have, respectively, 45 and 15 independent components and  satisfy the equations:
\bea
d{\cal S}_1^{abcd}&=&{\cal P}^a\wedge {\cal Y}^{cbd}+{\cal P}^a\wedge {\cal Y}^{dbc}+{\cal P}^b\wedge
   {\cal Y}^{cad}+{\cal P}^b\wedge {\cal Y}^{dac} \label{dS1} \\ &&-{\cal P}^c\wedge
   {\cal Y}^{adb}-{\cal P}^c\wedge {\cal Y}^{bda}-{\cal P}^d\wedge
   {\cal Y}^{acb}-{\cal P}^d\wedge {\cal Y}^{bca},\nn\\
d {\cal S}_2^{abcd}&=&4~{\cal Z}^{ab}\wedge
  {\cal Z}^{cd}+{\cal P}^a\wedge {\cal Y}^{bcd}-{\cal P}^b\wedge {\cal Y}^{acd}-{\cal P}^c\wedge
   {\cal Y}^{dab}+{\cal P}^d\wedge {\cal Y}^{cab}.\label{dS2}
\eea

 In the third extension the new non-vanishing commutators  are those with a total of four free indices. From  \bref{dS1},  \bref{dS2} it follows that they satisfy: \bea
\left[Z_{ab},~Z_{cd}\right]&=&4~ i ~S^2_{abcd},\\
\left[P_a,~Y_{bcd}\right]&=&i~(S^1_{acbd}-S^1_{adbc}) +~\frac{i}{3}(2S^2_{abcd}-S^2_{acdb}-S^2_{adbc}),
 \eea
where the new generators $S^1_{abcd}, ~S^2_{abcd}$ are assumed to have the full symmetries of the corresponding 1-form potentials.

The coset now will be written \bea
g&=&e^{iP_ax^a}e^{\frac{i}{2}Z_{ab}\theta^{ab}}
e^{\frac{i}{2}Y_{abc}\xi^{abc}}e^{\frac{i}{8}S^1_{abcd}\sigma_1^{abcd}}
e^{\frac{i}{8}S^2_{abcd}\sigma_2^{abcd}},\eea
where $\sigma_1^{abcd},~\sigma_2^{abcd}$ are new scalar
parameters having the symmetries of ${\cal S}_1^{abcd},~{\cal S}_2^{abcd}$.
After a long calculation we obtain the following explicit expressions for the new MC 1-forms:
\bea
{\cal S}_1^{abcd}&=&d\sigma_1^{abcd} -(dx^a\xi^{cbd}+dx^a\xi^{dbc}+dx^b\xi^{cad}+dx^b\xi^{dac})\\&&+dx^c\xi^{adb}+dx^c\xi^{bda}+dx^d\xi^{acb}+dx^d\xi^{bca}+\frac12 [x^ax^c(xdx)^{bd}+x^bx^d(xdx)^{ac}],\nn\\
{\cal S}_2^{abcd}&=&
d\sigma_2^{abcd}-(dx^a\xi^{bcd}-dx^b\xi^{acd}-dx^c\xi^{dab}+dx^d\xi^{cab})\\&&+2\theta^{ab}(d\theta^{cd}+(xdx)^{cd})-2\theta^{cd}(d\theta^{ab}+(xdx)^{ab}),\nn
 \eea
where we have used the notation $(xdx)^{ab}=(x^adx^b-x^bdx^a)$.

\subsection{Third order Lagrangian and equations of motion}
With the third order extensions, the particle Lagrangian becomes
 \be\label{lag3} \CL= m\sqrt{-\dot x_a^2}
+\frac12 f_{ab}{\cal Z}^{ab}+\frac12 f_{abc}{\cal Y}^{abc}+
\frac18 g_{abcd}{\cal S}_1^{abcd}+\frac18 h_{abcd}{\cal S}_2^{abcd},\ee
where  the new   tensor couplings $g_{abcd}(\tau)$, $h_{abcd}(\tau)$
   have the symmetries of ${\cal S}_1^{abcd},~{\cal S}_2^{abcd}$, respectively,
    and  together with the new group space coordinates $\sigma_1^{abcd}, \sigma_2^{abcd}$
are also treated as new dynamical variables.

This Lagrangian is invariant under the transformations found before
plus the following ones for the new variables
\bea \D_p \sigma^{abcd}_1&=&-\frac 32 (\epsilon^a x^b x^c
x^d+\epsilon^b x^a x^c x^d-\epsilon^c x^a x^b x^d-\epsilon^d x^a x^b
x^c),\\
\D_Z \sigma^{abcd}_1&=& 3(x^a x^d\epsilon^{bc}+x^a
x^c\epsilon^{bd}+x^b x^c\epsilon^{ad}+x^b x^d\epsilon^{ac}),\\
\D_Y \sigma^{abcd}_1&=&
x^a(\epsilon^{cbd}+\epsilon^{dbc})+x^b(\epsilon^{cad}+\epsilon^{dac})
-x^c(\epsilon^{adb}+\epsilon^{bda})-x^d(\epsilon^{acb}+\epsilon^{bca}),\\
\D_{S_1} \sigma^{abcd}_1&=&\epsilon^{abcd}_1 \eea
and
 \bea \D_p \sigma^{abcd}_2&=&(\epsilon^a x^b-\epsilon^b
x^a)\theta^{cd}-(\epsilon^c x^d-\epsilon^d x^c)\theta^{ab},\\
\D_Z \sigma^{abcd}_2&=&\epsilon^{ac}x^bx^d+ \epsilon^{bd}x^ax^c
-\epsilon^{ad}x^bx^c-\epsilon^{bc}x^ax^d-2\epsilon^{ab}\theta^{cd}+2\epsilon^{cd}\theta^{ab},\\
 \D_Y \sigma^{abcd}_2&=&
x^a\epsilon^{bcd}-x^b\epsilon^{acd}-x^c\epsilon^{dab}+x^d\epsilon^{cab},\\
\D_{S_2} \sigma^{abcd}_2&=&\epsilon^{abcd}_2. \eea

 The conserved quantities are written as \be
Q=\epsilon^a P_a+\frac12\epsilon^{bc}Z_{bc}+\frac12\epsilon^{abc}Y_{abc}
+\frac 18\epsilon^{abcd}_1 S^1_{abcd}+\frac 18\epsilon^{abcd}_2 S^2_{abcd} ,\ee
from which we obtain the Noether generators \bea
P_a&=&p_a-\frac12p_{ab}x^b+p_{bca}x^bx^c-\frac 34p^1_{abcd}
x^bx^cx^d+\frac 12 p^2_{abcd}x^b\theta^{cd}\\
 Z_{bc}&=&p_{bc}+x^d(2p_{dbc}-p_{bcd}-p_{cdb})\nn\\&&+\frac
 32(p^1_{abcd}-p^1_{acbd})x^a x^d-\frac
 12(p^2_{abcd}-p^2_{acbd})x^a x^d-p^2_{bcad}\theta^{ad},\\
Y_{abc}&=&p_{abc}+\frac 12(p^1_{abcd}-p^1_{acbd})x^d+
\frac13(2p^2_{dabc}-p^2_{dbca}-p^2_{dcab})x^d,\\
S^1_{abcd}&=&p_{abcd}^1,\quad S^2_{abcd}=p_{abcd}^2. \eea

The Euler-Lagrange  equations of motion can be reduced to \bref{theta2},   \bref{eomf3},  \bref{eomf2} and the following new equations:
\bea
\D {\sigma_1}^{abcd}&\to& \dot g_{abcd} =0,\label{g4}\\
\D  {\sigma_2}^{abcd}&\to& \dot h_{abcd} =0,\label{h4}\\
\D g_{abcd}&\to& \dot  {\sigma_1}^{abcd}= \dot x^a(\xi^{cbd}+\xi^{dbc})+ \dot x^b(\xi^{cad}+\xi^{dac})+ \dot x^c(\xi^{adb}+\xi^{bda})\\&&+\dot x^d(\xi^{acb}+\xi^{bca})\label{dotsig1}-\frac{1}{2}[
x^ax^c(x^b\dot x^d-x^d\dot x^b)+x^bx^d(x^a\dot x^c-x^c\dot x^a)],\nn\\
\D h_{abcd}&\to& \dot  {\sigma_2}^{abcd}=\dot x^a\xi^{bcd}-\dot x^b\xi^{acd}-\dot x^c\xi^{dab}+\dot x^d\xi^{cab}\label{dotsig2}\\&&-\T^{ab}(x^c\dot x^d-x^d\dot x^c)+\T^{cd}(x^a\dot x^b-x^b\dot x^a),\nn\\
\D \xi^{abc}&\to& \dot f_{abc}=-\dot x^d(g_{abcd}-g_{acbd})-\frac{\dot x^d}{3}( 2h_{dabc}-h_{dbca}-h_{dcab}),\label{theta3Ord3}\\
\D x^{a}&\to& -m\ddot x_a+f_{ab}\dot x^b=0, \label{eomxOrd3} \eea
where, in reducing    \bref{dotsig2},   \bref{eomxOrd3}, we have used  \bref{theta2},   \bref{eomf3},  \bref{eomf2}.
We should remark that, as with \bref{eomx}, the RHS of \bref{eomxOrd3} is not identically zero, but vanishes because of the other equations of motion. An extra term that vanishes because of \bref{eomf2} also appears on the RHS of \bref{theta2}.

  As in the previous levels, the last terms in \bref{dotsig1} allow
us to relate ${\sigma_1}^{abcd}$ to the third order moments (octupole) of the
current distribution. However,  $ \dot {\sigma_2}^{abcd}$ which
vanishes when $\T$ and $\xi$ vanish,   must be interpreted
differently: it arises from non-linear couplings of the current with the
quadrupole moment ($\dot x^a \xi^{cbd}$ terms) as well as of
$\T$ with  $\dot \T$ (using \bref{eomf2}, the last two terms in \bref{dotsig2} are $2\,\T^{ab}\dot{\T^{cd}}-2\,\T^{cd}\dot{\T^{ab}}$).

Integrating  \bref{g4},   \bref{h4},  \bref{theta3Ord3} and substituting in \bref{theta2}, we obtain the equation satisfied by $f_{ab}$:
\be
\dot f_{ab}=3(g^0_{cabd}-g^0_{cbad})x^c\dot x^d-(2h^0_{cdab}-h^0_{cabd}-h^0_{cbda})x^c\dot x^d+(-2f^0_{cab}+f^0_{abc}+f^0_{bca}) \dot x^c.
\ee
Writing $2x^c\dot x^d=\frac{d}{dt}(x^c x^d-2 \T^{cd})$, as follows from \bref{eomf2}, we can integrate this equation to get:
\bea
 f_{ab}&=&\frac34(g^0_{cabd}-g^0_{cbad}+g^0_{dabc}-g^0_{dbac})x^c x^d+\frac14(h^0_{cabd}-h^0_{cbad}+h^0_{dabc}-h^0_{dbac})x^c x^d\nn\\&&-2h^0_{abcd} \T^{cd}+(-2f^0_{cab}+f^0_{abc}+f^0_{bca}) x^c+f^0_{ab}.
 \eea

We observe that, when $h^0_{abcd} \ne 0$, the tensor $ f_{ab}$
depends on $\T^{ab}$ in addition to having terms quadratic in the
cartesian coordinates. Thus, only part of  $ f_{ab}$  can be derived
from a potential, and the interaction described by  $ f_{ab}$ can no
longer be interpreted as a  pure  electromagnetic field. The part of
$ f_{ab}$  that cannot be derived from a potential gives
   terms to the equation of motion that couple to the magnetic moment. Thus we write \bref{eomxOrd3} as
 \be m \ddot{x}_a+2h^0_{abcd}
\dot{x}^b\theta^{cd}-h^0_{abcd}x^b\dot{\T}^{cd}= e\, F_{ab}\dot{x}^b
\label{newEqMot3}, \ee where $F_{ab}$ represents an ordinary external
EM field,  now quadratic in the coordinates,  \be
F_{ab}=\frac34(g^0_{cabd}-g^0_{cbad}+g^0_{dabc}-g^0_{dbac})x^c
x^d+(-2f^0_{cab}+f^0_{abc}+f^0_{bca}) x^c+F^0_{ab}.\label{fab3}\ee  The
symmetries of  $g_{abcd}$ implies that  $F_{[ab,c]}=0$ and thus can
be derived from a potential. The terms depending on $h^0_{abcd}$  in
the equation of motion imply a damping effect due to the magnetic
moment. As the magnetic moment is determined by the position $x$,
this describes a back-reaction altering the time evolution of $x$.
Thus the dynamics of the motion now depends on the dynamics of the
new variables $\T^{ab}$. We can conjecture that at the next level
the tensor $ f_{ab}$ will contain terms depending on $\xi^{abc}$,
terms cubic in $x^{a}$ and terms with the product $\T^{ab}x^c$  and
thus the equation of motion for the coordinates will also couple to
the quadrupole  moment.

As we did with \bref{theta2}, writing \bref{theta3Ord3} as $df_{abc}
=-d x^d(g_{abcd}-g_{acbd})-d x^d( 2h_{dabc}-h_{dbca}-h_{dcab})/3$,
we can interpret the coefficients $g_{abcd}$ and $h_{abcd}$ as
giving the partial derivatives of $f_{abc}$, which, in turn,
determine the partial derivatives of $f_{ab}$.  Thus  $g_{abcd}$
and $h_{abcd}$ are, effectively, the second derivatives of
$f_{ab}$. Of course, the physical meaning of  these two
types of second derivatives is different.
The terms with $h_{abcd}$, leading to non-zero $f_{[ab,c]}$,
can be interpreted as magnetic sources. It is known that Maxwell's theory is
consistent with the existence of such sources.
However, we prefer to interpret these terms as introducing a coupling to the magnetic
moment ${\dot{\T}}^{ab}$ in the equation of motion for the position
variables.

The form \bref{newEqMot3} of the equation of motion, together with
the equations determining the evolution of the different moments
($\T^{ab}, \xi^{abc}, \sigma^{abcd}, \dots$), reinforce our
conclusion,  proposed  at the end of section 3, that the
physical system described here is a distribution of charges moving
consistently (including effects due to non-vanishing moments) in an
given EM field.  The description is approximated by a series
expansion of the field and the corresponding collection of moments
of the current distribution, successive levels in the extension
procedure giving higher approximations.

At the mathematical level, the coordinates of the extended group
space describe the degrees of freedom in the multipole  expansion of
the current distribution  and the induced interactions between them.
 And the  symmetry group
describes how changes in the coordinates and  multipole moments are
interrelated in order that a self-consistent  interpretation in terms of
moving charges in a   given external EM field,
including back-reaction terms, be possible.

\subsection{Fourth extension}

At the 4th level, the procedure gives 204 new  extensions which can be grouped as the components of 5 different 5th rank tensors with definite symmetries. The corresponding 1-form generators, denoted by the symbols ${\cal T}_i$, $i=1,...,5$ and having the symmetries indicated, satisfy:
\bea
d{\cal T}_1^{(abcd)e}&=&{\cal P}^{(a}\wedge {\cal S}_1^{bcd)e},\\
d {\cal T}_2^{(abc)(de)}&=&{\cal S}_1^{(abc)d}\wedge {\cal P}^{e}+{\cal S}_1^{(abc)e}\wedge {\cal P}^{d}-4({\cal Z}^{d(a}\wedge {\cal Y}^{bc)e}+{\cal Z}^{e(a}\wedge {\cal Y}^{bc)d})\nn\\ & &+\frac{4}{3}({\cal P}^{(a}\wedge {\cal S}_1^{bcd)e}+{\cal P}^{(a}\wedge {\cal S}_1^{bce)d}),\\
d {\cal T}_3^{(abc)[de]}&=&({\cal S}_1^{(abc)d}\wedge {\cal P}^e-{\cal S}_1^{(abc)e}\wedge {\cal P}^d+\frac85({\cal Z}^{d(a}\wedge {\cal Y}^{bc)e}-{\cal Z}^{e(a}\wedge {\cal Y}^{bc)d})\\&&\frac65({\cal S}_2^{adbe}\wedge {\cal P}^c-{\cal S}_2^{aebd}\wedge {\cal P}^c)_{(abc)}-\frac45({\cal P}^{(a}\wedge {\cal S}_1^{bcd)e}-{\cal P}^{(a}\wedge {\cal S}_1^{bce)d}),\nn\\\
d {\cal T}_4^{[abc][de]}&=&{\cal S}_2^{[abc]d}\wedge {\cal P}^e-{\cal S}_2^{[abc]e}\wedge {\cal P}^d+4({\cal Z}^{d[a}\wedge {\cal Y}^{bc]e}-{\cal Z}^{e[a}\wedge {\cal Y}^{bc]d})\\&&+\frac43({\cal P}^{[a}\wedge {\cal S}_2^{bcd]e}-{\cal P}^{[a}\wedge {\cal S}_2^{bce]d}+2{\cal Y}^{e[ab}\wedge{\cal Z}^{cd]}-2{\cal Y}^{d[ab}\wedge{\cal Z}^{ce]}),\nn\\
d {\cal T}_5^{[abcd]e}&=&{\cal P}^{[a}\wedge {\cal S}_2^{bcd]e}+2{\cal Y}^{e[ab}\wedge{\cal Z}^{cd]},
\eea
where the subscript $(abc)$ indicates the symmetry operation that must be applied to the expression in parenthesis. These  5th rank tensors have, respectively, (84, 60, 36, 20, 4) independent components.
We will not investigate further  the 4th level extensions.

 \section{Young tableau symmetries and possible 5th level extensions}

To understand the structure of the higher level extensions, it is helpful to discuss their symmetry properties in terms of Young Tableaux (YT).  The YT symmetries of all generators (MC 1-forms) up to level 4 are\footnote{The third level extensions have no particular YT symmetry, but can be expressed in terms of such tensors: $S_1^{(ab)(cd)}=YT_{31}^{(abc)d}+YT_{31}^{(abd)c}-YT_{31}^{(cda)b}-YT_{31}^{(cdb)a}, ~
S_2^{[ab][cd]}=YT_{211}^{[abc]d}-YT_{211}^{[abd]c}-YT_{211}^{[cda]b}+YT_{211}^{[cdb]a}$, where the indices on $YT$ tensors indicate the number of boxes in each row of the YT diagram. Similarly, for the fourth extension tensors ${\cal T}_i$ defined in subsection 4.3 to have the corresponding YT symmetry certain symmetry operations must be performed on each ${\cal T}_i$.} :

  $ {\tiny\Yvcentermath1{\cal P}^a   ~~  \yng(1),  ~{\cal Z}^{ab}   ~~  \yng(1,1), ~ {\cal Y}^{abc}   ~~  \yng(2,1),  ~{\cal S}_1^{abcd}   ~~  \yng(3,1), ~ {\cal S}_2^{abcd}   ~~  \yng(2,1,1)}$, \\
$~~ {\tiny\Yvcentermath1{\cal T}_1^{abcde}   ~~  \yng(4,1), ~ {\cal T}_2^{abcde}   ~~  \yng(3,2) ,~  {\cal T}_3^{abcde}   ~~  \yng(3,1,1), ~ {\cal T}_4^{abcde}   ~~  \yng(2,2,1), ~ {\cal T}_5^{abcde}   ~~  \yng(2,1,1,1)}$.

Note that completely symmetric YT do not appear, as the requirement that the exterior derivative of these 1-forms be given in terms of the wedge-product of lower generators implies at least one antisymmetry.
The 5th level generators (MC 1-forms) will have 6 indices and will
 transform as  the components of 6th rank tensors with the following possible (in 4 dimensions)  YT symmetries:

$ {\tiny\Yvcentermath1{\cal W}_1^{abcdef}   ~~  \yng(5,1), ~ {\cal W}_2^{abcdef}   ~~  \yng(4,2) ,~  {\cal W}_3^{abcdef}   ~~  \yng(4,1,1), ~ {\cal W}_4^{abcdef}   ~~  \yng(3,3)}$,\\  $ {\tiny\Yvcentermath1~ {\cal W}_5^{abcdef}   ~~  \yng(3,2,1) ,~  {\cal W}_6^{abcdef}   ~~  \yng(3,1,1,1), ~ {\cal W}_7^{abcdef}   ~~  \yng(2,2,2), ~ {\cal W}_8^{abcdef}   ~~  \yng(2,2,1,1)}$.

The exterior derivative of each ${\cal W}_i$ tensor will then be given in terms of wedge products of pairs of lower order generators whose YT can be combined to give the YT of  ${\cal W}_i$. For example, the YT $ {\tiny\Yvcentermath1  \yng(4,2) }$ can be obtained by multiplying the following pairs of YT:
$ {\tiny\Yvcentermath1 \yng(4,1)\otimes \yng(1), ~~\yng(3,2)\otimes \yng(1),  ~~\yng(3,1)\otimes \yng(1,1), ~~ \yng(2,1)\otimes \yng(2,1)}$.

Thus, we expect $d {\cal W}_2^{(abcd)(ef)}$ to be given  as a linear combination of the following 4 terms:  $ {\cal T}_1^{(abcd)(e}\wedge {\cal P}^{f)}$,  ~$ {\cal P}^{(a}\wedge {\cal T}_2^{bcd)(ef)}$,~ $ {\cal Z}^{e(a}\wedge {\cal S}_1^{bcd)f}+ {\cal Z}^{f(a}\wedge {\cal S}_1^{bcd)e}$ and the $(abcd)(ef)$ part of $ {\cal Y}^{abe}\wedge {\cal Y}^{cdf}$, which is, however, identically zero. Requiring the exterior derivative of this linear combination to be zero, we determine the unknown coefficients (up to an overall constant factor) and thus the equation defining ${\cal W}_2$:
\be
d {\cal W}_2^{(abcd)(ef)}=10~{\cal T}_1^{(abcd)(e}\wedge {\cal P}^{f)}-3{\cal P}^{(a}\wedge {\cal T}_2^{bcd)(ef)}+3({\cal Z}^{e(a}\wedge {\cal S}_1^{bcd)f}+ {\cal Z}^{f(a}\wedge {\cal S}_1^{bcd)e}).
\ee
Not all possible YT symmetries with a given number of indices are present: already at level 3,
the symmetry ${\tiny\Yvcentermath1 \yng(2,2)}$ does not appear. Conversely, some YT symmetries may appear more than once, i.e., the  linear combination of possible 2 forms with a given symmetry may not be unique. Due to computer memory limitations, we have done the 5th-level calculations in 3 dimensions and found that the generator  ${\cal W}_7$ does not appear, while the generators  ${\cal W}_3$ and  ${\cal W}_5$ appear with multiplicities 2 and 3, respectively (the generators  ${\cal W}_6$ and  ${\cal W}_8$, being antisymmetric in 4 indices, cannot exist).

We observe that in the Lagrangian, the coupling to the most
symmetric generator (say $k_{(abcd)e}$ corresponding to  ${\cal
T}_1^{(abcd)e}$) will only contribute a term depending on the
coordinates to $f_{ab}$. Thus, restricting attention to these most
symmetric extensions, $f_{ab}$  will satisfy  $f_{[ab,c]}=0$,\footnote{Consider the Taylor
expansion of $f_{ab}$ about the origin and let  $q_{[ab](cde)}$    denote  the  coefficients of the cubic term. When  $f_{ab}$  satisfies $f_{[ab,c]}=0$,  these coefficients  must satisfy  $q_{[abc](de)}=0$. These conditions imply that the  $q_{[ab](cde)}$ coefficients have   the same YT symmetry as that of the coefficients $k_{(abcd)e}$, which satisfy $k_{(abcde)}=0$, and therefore the $k_{(abcd)e}$ coefficients determine the part  cubic in the coordinates of an antisymmetric tensor  $f_{ab}$  satisfying  $f_{[ab,c]}=0$.}  so
that an interpretation in terms of an ordinary electromagnetic field
will be always possible.

 We observe  that these coefficients associated to the  most
symmetric extensions are in one to one correspondence with the zero
forms  used by Vasliliev to describe the ``unfolded dynamics" of the
Maxwell equations \cite{Vasiliev:2005zu}, see also
\cite{Skvortsov:2008vs}.

 \section{Summary and discussion}

We have studied in detail the structure and the particle dynamics of the
infinite sequence of extensions of the Poincar{\'e} algebra outlined in
\cite{Bonanos:2008kr}. We have seen that the generators of the
non-central extensions belong to tensor
representations of the Lorentz group of  increasingly higher
 rank. We can associate one or more Young tableaux to every extension. Although we have done the calculations in four dimensions, the extensions found exist in any dimension where their symmetry is allowed.  We conjecture that the complete set of extensions constitutes an infinite Lie algebra. 
 
We do not have a precise mathematical interpretation for  this
infinite algebra, 
but we note that its generator content
is organized in levels like the Lorentzian Kac-Moody
algebras, that are conjectured to be a symmetry of supergravity
\cite{West:2001as}, \cite{Damour:2002cu}. It is not completely
unnatural that there might be some connection between the two
structures, despite the fact that the fields in the Lorentzian
Kac-Moody algebras at level zero include the graviton.
Following this direction we have studied analogies with the
representations of the over-extension of the $G_2$,
$G_2^{++}$\footnote{  d=5 N=2 pure supergravity using $G_2^{++}$ was
studied in \cite{Mizoguchi:2005zf}.}
 algebras with respect to $A_3$\footnote{We have used the computer
program SimpLie \cite{Teake} to study the level structure of the
corresponding representations.}. If we disregard the level zero, at
level one there is a vector corresponding to $P_a$, at level two an
antisymmetric two tensor $Z_{ab}$, at level three a mixed generator that
corresponds to our $Y_{abc}$, but at level four only one object that
corresponds to $S^2_{abcd}$ exists; the generator $S^1_{abcd}$ does
not appear. At level 5 only $T^4_{abcde}, T^5_{abcde}$ appear. At
level 6 there appears only $W^5_{abcdef}, W^8_{abcdef}$. Therefore
we can conclude that only some of the extensions of the infinite
sequence of Poincar{\'e} algebras we found appear also at non-zero
levels of $G_2^{++}$. We do not know if one can find an infinite
algebra that encompasses all the possible Poincar{\'e} extensions.

In order to understand the physical significance of this infinite
sequence of extensions of the Poincar{\'e} group, we have constructed an
 invariant Lagrangian that depends linearly on the extensions by
introducing tensor  coupling ``constants" that we consider as new dynamical variables.
 The physical system described by this Lagrangian is a distribution of charged particles moving in an external electromagnetic field. The description is approximate: the particles are described collectively by their multipole moments about the world line of their center of mass and the field  by its Taylor expansion about the same line. New terms in the approximation series appear with every extension. The multipoles can be considered as Goldstone bosons. The higher extensions give back-reaction terms describing the effect of the moments on the world line.

We think that sequential extensions of space-time
groups including odd generators,
 using the same methods as in this paper, might be useful in constructing 
 theories containing fermions.
We hope to address this point in the future.

\section*{Acknowledgments}
We acknowledge Adolfo Azcarraga, Andrea Barducci, Roberto
Casalbuoni, Jaume Gomis, Machiko Hatsuda,  Satoshi Iso, Yoshihisa
Kitazawa, Kiyoshi Kamimura, Jerzy Lukierski, Shun'ya Mizoguchi,
Teake Nutma, Norisuke Sakai and George Savvidy  for discussions.
This work has been supported by MCYT FPA 2007-66665, CIRIT GC
2005SGR-00564, Spanish Consolider-Ingenio 2010 Programme CPAN
(CSD2007-00042).

We are also grateful to the referees for urging us to provide  a physical interpretation for the extended group coordinates.


\begin{thebibliography}{99}

\bibitem{Bonanos:2008kr}
  S.~Bonanos and J.~Gomis,
  ``A note on the Chevalley-Eilenberg Cohomology for the Galilei and Poincar{\'e}
  Algebras,'' J. Phys. A: Math. Theor. {\bf 42} (2009) 145206
 [arXiv:hep-th/0808.2243].

\bibitem{levyleblond69}
J.-M. L\'evy-Leblond, ``Group-Theoretical Foundations of Classical
Mechanics: The Lagrangian Gauge Problem," Comm. Math. Phys. {\bf 12}
(1969) 64.

\bibitem{Aldaya:1985va}
  V.~Aldaya, J.~A.~de Azcarraga and R.~W.~Tucker,
  ``Group Cohomology Approach To The Quantum Equivalence Principle,''
  J. Geom. Phys. {\bf 3}, 303 (1986).


\bibitem{olmo1}
M.~ Gadella, M.~A.~Martin, L.~Nieto and M.~del Olmo `` The
Stratonovich-Weyl correspondence for one dimensional kinematical
groups," J. Math. Phys. {\bf 32} (1991) 1182.

\bibitem{Cangemi:1992bj}
  D.~Cangemi and R.~Jackiw,
  ``Gauge Invariant Formulations Of Lineal Gravity,''
  Phys.\ Rev.\ Lett.\  {\bf 69} (1992) 233
  [arXiv:hep-th/9203056].

\bibitem{Galperin:1987wc}
  A.~S.~Galperin, E.~A.~Ivanov, V.~I.~Ogievetsky and E.~Sokatchev,
  ``Gauge Field Geometry from Complex and Harmonic Analyticities. I. K\"{a}hler and
  Self-Dual Yang-Mills Cases,''
  Ann. Phys.\  {\bf 185} (1988) 1.


\bibitem{Galperin:1987wb}
  A.~S.~Galperin, E.~A.~Ivanov, V.~I.~Ogievetsky and E.~Sokatchev,
  ``Gauge Field Geometry from Complex and Harmonic Analyticities. II. Hyper-K\"{a}hler
  Case,''
  Ann. Phys.\  {\bf 185} (1988) 22.


\bibitem{galindo}
A. ~Galindo ``Lie algebra extensions of the Poincar\'e Algebra,''
  J.\ Math.\ Phys.\  {\bf 8} (1967) 768.


\bibitem{Schrader:1972zd}
  R.~Schrader,
  ``The Maxwell Group and the Quantum Theory of Particles in Classical Homogeneous Electromagnetic Fields,''
  Fortsch.\ Phys.\  {\bf 20} (1972) 701.

\bibitem{Beckers:1983gp}
  J.~Beckers and V.~Hussin,
  ``Minimal Electromagnetic Coupling Schemes. Ii. Relativistic And
  Nonrelativistic Maxwell Groups,''
  J.\ Math.\ Phys.\  {\bf 24} (1983) 1295.


\bibitem{Bacry:1970ye}
  H.~Bacry, P.~Combe and J.~L.~Richard,
  ``Group-theoretical analysis of elementary particles in an external
  electromagnetic field. 1. the relativistic particle in a constant and uniform
  field,''
  Nuovo Cim.\  A {\bf 67} (1970) 267.

\bibitem{larsson}
T.~A.~Larsson,
``Extensions of diffeomorphism and current algebras,"
arXiv:math-ph/0002016


\bibitem{azcarragabook}
  J.~A.~de Azcarraga and J.~M.~Izquierdo,
\emph{``Lie groups, Lie algebras, cohomology and some applications
in physics.''} Cambridge Univ. Press, 1995.



\bibitem{bonanos} S. Bonanos,  ``Exterior Differential Calculus",
Mathematica Enhancement,
\url{http://www.inp.demokritos.gr/~sbonano/EDC/}



\bibitem{West:2001as}
  P.~C.~West,
  ``E(11) and M theory,''
  Class.\ Quant.\ Grav.\  {\bf 18} (2001) 4443
  [arXiv:hep-th/0104081].

\bibitem{Damour:2002cu}
  T.~Damour, M.~Henneaux and H.~Nicolai,
  ``E(10) and a `small tension expansion' of M theory,''
  Phys.\ Rev.\ Lett.\  {\bf 89}, 221601 (2002)
  [arXiv:hep-th/0207267].

T.~Damour, M.~Henneaux and H.~Nicolai,
  ``Cosmological billiards,''
  Class.\ Quant.\ Grav.\  {\bf 20}, R145 (2003)
  [arXiv:hep-th/0212256].




\bibitem{Coleman}
  S.~R.~Coleman, J.~Wess and B.~Zumino,
``Structure of phenomenological Lagrangians. I,''
  Phys.\ Rev.\  {\bf 177} (1969) 2239;

  C.~G.~.~Callan, S.~R.~Coleman, J.~Wess and B.~Zumino,
  ``Structure of phenomenological Lagrangians. II,''
  Phys.\ Rev.\  {\bf 177} (1969) 2247.

\bibitem{Gomis:2006xw}
  J.~Gomis, K.~Kamimura and P.~C.~West,
  ``The construction of brane and superbrane actions using non-linear
  realisations,''
  Class.\ Quant.\ Grav.\  {\bf 23} (2006) 7369
  [arXiv:hep-th/0607057].
  
\bibitem{Vasiliev:2005zu}
  M.~A.~Vasiliev,
  Int.\ J.\ Geom.\ Meth.\ Mod.\ Phys.\  {\bf 3} (2006) 37
  [arXiv:hep-th/0504090].


\bibitem{Skvortsov:2008vs}
  E.~D.~Skvortsov,
  JHEP {\bf 0807} (2008) 004
  [arXiv:hep-th/0801.2268].



\bibitem{Mizoguchi:2005zf}
  S.~Mizoguchi, K.~Mohri and Y.~Yamada,
  ``Five-dimensional supergravity and hyperbolic Kac-Moody algebra G(2)(H),''
  Class.\ Quant.\ Grav.\  {\bf 23} (2006) 3181
  [arXiv:hep-th/0512092].



\bibitem{Teake} T. Nutma
``SimpLie: a simple program for Lie algebras",
\url{http://code.google.com/p/simplie/}

\end{thebibliography}
\end{document}